# How important is choice of the scaling factor in standardizing citations?[1]

*Giovanni Abramo[a,b,*], Tindaro Cicero[b], Ciriaco Andrea D'Angelo[b]*

[a] Institute for System Analysis and Computer Science (IASI-CNR)
National Research Council of Italy

[b] Laboratory for Studies of Research and Technology Transfer
School of Engineering, Department of Management
University of Rome "Tor Vergata"

**Abstract**

Because of the variations in citation behavior across research fields, appropriate standardization must be applied as part of any bibliometric analysis of the productivity of individual scientists and research organizations. Such standardization involves scaling by some factor that characterizes the distribution of the citations of articles from the same year and subject category. In this work we conduct an analysis of the sensitivity of researchers' productivity rankings to the scaling factor chosen to standardize their citations. To do this we first prepare the productivity rankings for all researchers (more than 30,000) operating in the hard sciences in Italy, over the period 2004-2008. We then measure the shifts in rankings caused by adopting scaling factors other than the particular factor that seems more effective for comparing the impact of publications in different fields: the citation average of the distribution of cited-only publications.

* **Corresponding author**: Dipartimento di Ingegneria dell'Impresa, Università degli Studi di Roma "Tor Vergata", Via del Politecnico 1, 00133 Rome - ITALY, tel/fax +39 06 72597362, giovanni.abramo@uniroma2.it

# 1. Introduction

Field-standardization of citations is now common practice for any serious bibliometric analysis, applied to comparative measurement of research performance of individuals, entire organizations, departments, or other units. This is necessary because of the different citation behavior of researchers in various fields. A number of studies have shown that there is generally a different time distribution of citations across fields (Hurt, 1987; Peters and Van Raan, 1994; Redner, 1998; Gupta et al., 2005; Peterson et al., 2010; Stringer et al., 2010; Vieira and Gomes, 2010). In schematic terms, the number of citations observed at time *t* for an article in mathematics is different from the number observed at the same time for an article of the same quality in physics, published in the same year. To make citations comparable for articles that belong to different fields, bibliometricians standardize citations by applying a scaling factor. Failure to carry out such field standardization can cause notable distortions in measures of performance, as demonstrated by various studies (Abramo and D'Angelo, 2007; Lundberg, 2007; Leydersdoff, 2011).

    Standardization involves first classifying each article according to its subject category and then scaling the citations. The scaling is carried out by multiplying the citations of each article by a factor that characterizes the distribution of the citations of all articles from the same year and subject category (for example the inverse of the median or mean). In actual practice, bibliometricians adopt different scaling factors. The well-known "crown indicator", originated by the Leiden University CWTS, scales the citations of a given publication set with respect to mean of the category distribution (Moed et al., 1995). The Karolinska Institute's "field normalized citation score" also uses the mean as scaling factor, applied to the citations for each publication (Rehn et al., 2007). Vinkler (1997) in his Relative Subfield Citedness (Rw) (where W refers to " world") indicator, relates the number of citations obtained by the set of papers evaluated to the number of citations received by a same number of papers published in journals dedicated to the respective discipline, field or subfield. The current authors introduced the "Scientific Strength" indicator. For this performance indicator, they originally standardized citations by the mean (Abramo and D'Angelo, 2011) but recently, observing the strong skewness of the citation distributions, have switched to the median of the distribution (Abramo et al., 2011). A different overall approach is seen in the "relative impact index" indicator, developed by the Swiss Federal Government's Centre for Science and Technology Studies and reported in the Bibliometric handbook for Karolinska Institutet[2]. Here, the citation count is fractionalized with regard to the length of the reference list. Other citation indicators take into account the highly skewness of citation distributions, rating each publication in terms of its percentile in the citation distribution (Leydersdoff et al., 2011; Bornmann and Mutz, 2010). Few scholars have carried out studies aimed at identifying the most appropriate scaling factor. Radicchi et al. (2008) showed that citations distributions from 20 different disciplines and years could be rescaled on a universal curve, by applying the scaling factor of average number of citations per article. Following up this work, Radicchi and Castellano (2011) later provided a deeper study of the fields exclusive to physics, and confirmed that "when a rescaling procedure by the average is used, it is possible to compare impartially articles across years and fields" and added that "the median is less sensitive to possible extreme

---

[2] http://kib.ki.se/sites/kib.ki.se/files/Bibliometric_indicators_definitions_1.0.pdf. Last accessed on Apr. 13, 2012. The web reference to the CEST provided by the Karolinska Institutet is no longer accessible.

events such as the presence of highly cited papers, but dividing the raw number of cites by the median value leads to less fair comparisons and only for sufficiently old publications". These empirical analyses refer to specific disciplines, and the extension of the results to other disciplines is not so readily assumed. Albarran et al. (2011) and Waltman et al. (2012), analyzing a much larger dataset of publications, confirmed that the results hold for many but not for all scientific fields. Recently, Radicchi and Castellano (2012), expanding the dataset for their analysis (about 4,000,000 documents published in 6 distinct years in 8,304 scientific journals), introduced a simple mapping able to transform the citations distribution within a specific field into a universal power law, which depends on two parameters. Each of them is specific of a field (i.e. subject category), but for the vast majority of subject categories, the power law exponent is constant. The only subject categories for which the transformation is not a power-law function are hybrid, such as Multidisciplinary sciences, or not well defined such as Engineering, petroleum or Biodiversity conservation. In contrast, Lundberg (2007) suggested that due to the strong skewness of distributions of citations, it was preferable to use the median or geometric mean to scale citations, but he then demonstrated that the "item oriented field normalized logarithm-based citation z-score average" (or citation z-score) was still better.

All these studies, intended to support the choice of the most effective scaling factor for evaluation exercises, suffer from the conditions surrounding the tests, which have not simulated the typical practices of an evaluation exercise. Recently Abramo et al. (2012) overcame this limitation by simulating the terms of reference of a typical national research assessment exercise. With reference to all Italian universities' publications in two different years, they compared the effectiveness of six different methods of standardizing citations for all subject categories in the hard sciences, and concluded that the citations average seems the most effective scaling factor, when the average is based only on the publications actually cited.

Observing that different practitioners adopt different methods, in this work we propose to conduct an analysis of the sensitivity of individual researchers' productivity rankings to the scaling factor chosen to standardize citations. The reference context for the study is the Italian university system, limited to the disciplines where scientific performance can be evaluated by means of bibliometric techniques, meaning the hard sciences. For each standardization mode we calculate the performance rankings for all researchers belonging to these science disciplines over the period 2004-2008. In light of the findings from the work cited above (Abramo et al., 2012), we take the performance rankings derived from standardization by citation average of cited-only publications as benchmark for our analysis. Finally, we measure shifts in rankings from the benchmark, caused by adopting different scaling factors. To the best of our knowledge, the literature does not offer any similar studies that compare and evaluate the results obtained from different scaling factors.

In the next section of the paper we illustrate the methodology for measurement of individual research performance, the reference dataset and the different scaling factors adopted. In the third section we compare the performance rankings obtained from the application of different scaling factors to the study population. In the final section we comment on the results and draw conclusions.

## 2. Methodology and dataset

Research activity is a production process in which the inputs consist of human, tangible (scientific instruments, materials, etc.), and intangible (accumulated knowledge, social networks, etc.) resources; and where output, i.e. the new knowledge, has a complex character of both tangible nature (publications, patents, conference presentations, databases, protocols, etc.), and intangible nature (tacit knowledge, consulting activity, etc.). The new-knowledge production function has therefore a multi-input and multi-output character. The principal efficiency indicator of any production system is labor productivity. When measuring labor productivity, if there are differences in the production factors available to each scientist then one should normalize by them. Unfortunately, in Italy relevant data are not available at individual level. The first assumption then, is that resources available to researchers within the same field of observation are the same. The second assumption is that the hours devoted to research are more or less the same for all researchers. These assumptions are fairly well satisfied in the Italian higher education system, which is mostly public and not competitive. Up to 2009, the core funding by government was input oriented, meaning that it was distributed to universities in a manner intended to satisfy the needs for resources of each and all, in function of their size and activities. Furthermore, the time to devote to education is established by law.

In comparing productivity of individual researchers we consider only the following outputs: articles, article reviews and conference proceedings indexed in Web of Science (WoS) of Thomson Reuters. For the fields of analysis is limited to the hard sciences, this approximation is widely accepted in the literature. In fact, the choice of excluding other recognized outputs (letters, authored and edited books, technical reports, patents, prototypes, etc.) has clear empirical support (Moed et al., 2004). These forms of output are often followed by journal or conference publications that describe the content in the broad scientific arena, so the analysis of publications alone actually avoids a potential double counting.

The publications that embed the new knowledge have different values, depending on their impact on scientific advancement. As proxy of impact, we adopt the number of citations for the researchers' publications in a given period of observation.

Because the intensity of publications varies across fields, in order to avoid distortions in productivity rankings, comparisons among researchers should only be made between those working in the same field. A prerequisite of any performance assessment free of distortions is thus a classification of each researcher in one and only one field. In fact, in the Italian university system each researcher is classified in only one field. This feature of the Italian higher education system is unique in the world. In the hard sciences, there are 205 such fields (named scientific disciplinary sectors, SDSs[3]), grouped into nine disciplines (named university disciplinary areas, UDAs[4]). However, comparing researchers' performance by field is not enough to avoid distortions in rankings. This is because citation behavior also varies across fields, and it has been shown that it is not unlikely that researchers belonging to a particular scientific field may also publish outside that field (a typical example is statisticians, who may apply theory to medicine, physics, social sciences, etc.). For this reason, citations for each publication need to be standardized with respect to a scaling factor, such as the mean or the median for the distribution of citations for all the publications of the same year and the same subject category[5]. Furthermore, research projects frequently involve a team of researchers, demonstrated in the co-authorship of publications. More accurate productivity measures should then account for the fractional contributions of scientists to the outputs and, in the life sciences, for the position of co-authors in the list. Abramo and D'Angelo (2011) offer an example of application of the above methodology, to measure individual performance in Italian universities.

---

[3] The complete list is accessible on http://attiministeriali.miur.it/UserFiles/115.htm. Last accessed on Apr. 13, 2012.
[4] Mathematics and computer sciences, Physics, Chemistry, Earth sciences, Biology, Medicine; Agricultural and veterinary sciences, Civil engineering, Industrial and information engineering

The bibliometric dataset used in our analysis is extracted from the Italian Observatory of Public Research (ORP)[6], a database developed and maintained by the authors and derived under license from the Thomson Reuters WoS. Beginning from the raw data of the WoS and the database of research staff at all Italian universities, made available from the Ministry of Education, Universities and Research[7], and applying a complex algorithm for reconciliation of the author's affiliation and disambiguation of the true identity of the authors, each publication is attributed to the scientist that produced it (D'Angelo et al., 2011).

In this study we consider the authors in continuous faculty role over the period 2004-2008 and their relative scientific production, as indexed in the WoS over the same period. Citations are counted as of June 30, 2009. To ensure the representativity of publications as proxy of the research output, we consider only the scientists belonging to the 183 SDSs where at least 50% of researchers produced at least one publication in the period. The dataset is thus composed of 30,739 academic scientists and their respective publications, distributed as in Table 1.

| UDA | Research staff | Publications[8] | SDSs | Universities |
|---|---|---|---|---|
| Mathematics and computer sciences | 2,850 | 12,915 | 9 | 61 |
| Physics | 2,287 | 22,308 | 8 | 60 |
| Chemistry | 2,831 | 23,904 | 11 | 58 |
| Earth sciences | 1,100 | 4,373 | 12 | 48 |
| Biology | 4,480 | 26,525 | 19 | 63 |
| Medicine | 9,417 | 48,111 | 47 | 55 |
| Agricultural and veterinary sciences | 2,449 | 9,795 | 28 | 46 |
| Civil engineering | 1,127 | 3,834 | 7 | 47 |
| Industrial and information engineering | 4,198 | 30,427 | 42 | 63 |
| Total | 30,739 | 161,574 | 183 | 72 |

*Table 1: Research staff, publications, universities and SDSs of the Italian academic system, per UDA; data 2004-2008*

The performance of a single researcher, named Scientific Strength (SS), is calculated as[9]:

---

[5] The subject category of a publication is the same as the category of the journal where it is published; for publications in multidisciplinary journals the scaling factor is calculated as the average of the standardized values for each subject category.
[6] www.orp.researchvalue.it (last accessed on Apr. 13, 2012)
[7] See http://cercauniversita.cineca.it/php5/docenti/cerca.php
[8] The column total does not correspond to the sum of values per UDA because multiple counts occur when publications belong to more than one UDA.
[9] Because we are concerned with comparisons of rankings rather than with performance ranking per se, we do not calculate the much more complicated Fractional Scientific Strength, which involves the further precision of the contributions from each co-author to the publication.

$$SS = \sum_{i=1}^{N} AII_i$$

Where:

$AII_i$ = standardized citations, received by publication *i* of the researcher;
N = number of publications of the researcher in the period of observation.

To standardize the citations we use three different scaling factors, giving three alternative versions of the Article Impact Index (AII). For a general publication, with *c* indicating the citations received, we have:

i. $AII_a = \frac{c}{a}$
ii. $AII_{m0} = \frac{c}{m0}$
iii. $AII_{a0} = \frac{c}{a0}$

Where:

*a* = average of the distribution of citations received by all Italian publications of the same year and subject category;
*a0* = average of the distribution of citations received by the cited Italian publications of the same year and subject category (publications with no citations are not considered);
*m0* = median of the distribution of citations received by the Italian cited-only publications of the same year and subject category.

Standardization to the median of the distribution of citations received by all Italian publications (including not cited ones) is not considered, since there are a high number of SDSs with median of nil.

Because of the different intensity of publications across scientific fields, the comparison of individual research performance will be done at the level of SDS. Therefore we calculate researcher performance rankings for each standardization mode in each SDS. In addition, we calculate rankings by two other standardization methods. The first is by the absolute value of the individual's non-standardized citations, with the performance of each researcher ranked according to the sum of the citations of his/her publications over the period considered. The second ranking is obtained from the citation percentile rank of each publication in its respective subject category, with performance given by the sum of the percentile ranks of all the researcher's publications. Thompson (1993) warns that percentile ranks should not be added or averaged, because percentile is a numeral that does not represent equal-interval measurement. On the opposite, Leydersdoff et al. (2011) consider percentile ranks an improvement of averages-based indicators mainly because using non-parametric statistics one can abstract from the shape of the distribution of citations over papers. Comparing such sum rankings against benchmark permits an indication of the differences between these measures.

## 3. Results and analysis

To measure the shifts from benchmark we carry out a four step process. First is the standardization of citations for each publication in the dataset, in each SDS and for each standardization mode. Second is the calculation of rankings of individual performance by the researchers, in their SDSs. The third step is the overall measurement of the shifts in rankings with respect to the benchmark. Finally, we focus analysis on measurement of shifts for the top 25% and the bottom 25% of performers.

As a first step, we standardize the citations of each publication of our dataset by the scaling factors proposed and calculate the relative AII. As an example, Table 2 illustrates the calculation of AII values for the 2004 publications with more than 500 citations. For multi-category publications (e.g. ID=3), the value is given by the average of AII from all categories.

There are 13 publications identified: we observe the manner in which change in the choice of citation scaling factor brings variation in their relative rankings. For example, publication ID=2, which is second for absolute number of citations, results in a six-way tie for first place among the group, when ranking is by percentile in the subject category. It holds 8$^{th}$ position among the group when standardized to the average, 3$^{rd}$ position with standardization to median, and repeats at 8$^{th}$ using average of citations for only cited publications. Publication ID=1 is in first position no matter what type of standardization is used, while publications ID=4, 6, 10, 11 and 12 hold constant rankings if we exclude the simple performance measurement by number of citations.

| Pub_ID | Subject category* | Citations | Percentile | $AII_a$ | $AII_{m0}$ | $AII_{a0}$ |
|---|---|---|---|---|---|---|
| 1 | UI | 3,445 (1) | 100.00 (1) | 229.35 (1) | 430.63 (1) | 182.17 (1) |
| 2 | PY | 1,259 (2) | 100.00 (1) | 20.54 (8) | 78.69 (3) | 17.90 (8) |
| 3 | DB; KM | 1,056 (3) | 100.00 (1) | 78.22 (2) | 118.80 (2) | 65.13 (2) |
| 4 | PY | 759 (4) | 99.61 (9) | 12.38 (9) | 47.44 (9) | 10.79 (9) |
| 5 | RU | 692 (5) | 100.00 (1) | 43.61 (3) | 69.20 (4) | 39.15 (3) |
| 6 | PY | 682 (6) | 99.22 (10) | 11.13 (10) | 42.63 (10) | 9.70 (10) |
| 7 | DM | 613 (7) | 100 (1) | 34.61 (5) | 61.30 (5) | 33.08 (4) |
| 8 | DM | 610 (8) | 99.94 (8) | 34.44 (6) | 61.00 (6) | 32.92 (5) |
| 9 | BU | 592 (9) | 100 (1) | 36.98 (4) | 53.82 (7) | 29.23 (6) |
| 10 | PY | 561 (10) | 98.83 (11) | 9.15 (11) | 35.06 (11) | 7.98 (11) |
| 11 | PY | 534 (11) | 98.44 (12) | 8.71 (12) | 33.38 (12) | 7.59 (12) |
| 12 | PY | 517 (12) | 98.05 (13) | 8.44 (13) | 32.31 (13) | 7.35 (13) |
| 13 | RU | 504 (13) | 99.94 (7) | 31.76 (7) | 50.40 (8) | 28.51 (7) |

*Table 2: Subset of 2004 publications with more than 500 citations: values of citations; percentile rank and Article Impact Index (rank within the "over 500" subset in parenthesis)*
\* UI=physics, multidisciplinary; PY=medicine, general and internal; DB=biotechnology and applied microbiology; KM=genetics and heredity; RU=neurosciences; DM=oncology; BU=astronomy and astrophysics

After applying each scaling factor to standardize the citations received for all publications in the dataset, we apply indicator SS to measure the individual performance of Italian academic staff, in each SDS. We thus construct five scenarios, one for each indicator of individual SS performance, as illustrated in Table 3.

|   | Indicator | Description |
|---|---|---|
| 1 | $SS_{cit}$ | Scientific strength derived by the sum of citations |
| 2 | $SS_{perc}$ | Scientific strength derived by the sum of citations percentile |
| 3 | $SS_a$ | Scientific strength derived by the sum of $AII_a$ |
| 4 | $SS_{m0}$ | Scientific strength derived by the sum of $AII_{m0}$ |
| 5 | $SS_{a0}$ | Scientific strength derived by the sum of $AII_{a0}$ |

*Table 3: Modes of SS indicator used to evaluate individual researcher performance*

To remedy for the different sizes of SDSs, the rankings are prepared by percentile rank. Next we analyze the rank variations of researchers' performance with variation of the methodology for standardizing citations.

As an example, Table 4 shows the rankings lists for the scientists in the SDS AGR/10 (Rural construction and environmental land management) of UDA Agricultural and veterinary sciences, for each scenario. The rankings lists do not include researchers with nil SS (performance value independent of standardization). We observe that, for this SDS, the variations in percentile ranking are not particularly noticeable. Researcher_ID=1 is in the first percentile under any methodology of standardization, with the exception of $SS_{perc}$. Researcher #2 ranks among the top 10% with application of indicator $SS_{a0}$ or $SS_{cit}$, while any other indicator causes a lower rank, dropping as far as percentile 68.75, with $SS_a$.

The last line of the table shows the Spearman coefficient of correlation between the rankings from each scenario and the benchmark $SS_{a0}$ scenario. The coefficient permits us to measure the distance between the rankings and obtain a first measure of the distance incurred by using an impact indicator different from the benchmark reference.

In this SDS, the highest coefficient of correlation is obtained with the scenario of rankings from non-standardized citations ($\rho=+0.980$). Still, in general, the coefficients of correlation are extremely high: the lowest value is seen for the scenario with $SS_{perc}$ ($\rho=+0.850$).

| Researcher_ID | $SS_{cit}$ | $SS_{perc}$ | $SS_a$ | $SS_{m0}$ | $SS_{a0}$ |
|---|---|---|---|---|---|
| 1 | 100.00 | 93.75 | 100.00 | 100.00 | 100.00 |
| 2 | 93.75 | 75.00 | 68.75 | 81.25 | 93.75 |
| 3 | 87.50 | 87.50 | 87.50 | 93.75 | 87.50 |
| 4 | 87.50 | 87.50 | 87.50 | 93.75 | 87.50 |
| 5 | 75.00 | 68.75 | 75.00 | 75.00 | 75.00 |
| 6 | 62.50 | 100.00 | 93.75 | 68.75 | 68.75 |
| 7 | 68.75 | 43.75 | 62.50 | 62.50 | 62.50 |
| 8 | 56.25 | 62.50 | 56.25 | 56.25 | 56.25 |
| 9 | 56.25 | 25.00 | 37.50 | 37.50 | 50.00 |
| 10 | 56.25 | 25.00 | 37.50 | 37.50 | 50.00 |
| 11 | 56.25 | 50.00 | 43.75 | 43.75 | 37.50 |
| 12 | 31.25 | 56.25 | 50.00 | 50.00 | 31.25 |
| 13 | 18.75 | 37.50 | 25.00 | 25.00 | 25.00 |
| 14 | 25.00 | 0.00 | 12.50 | 18.75 | 18.75 |
| 15 | 18.75 | 31.25 | 18.75 | 18.75 | 12.50 |
| 16 | 18.75 | 12.50 | 6.25 | 6.25 | 6.25 |
| 17 | 18.75 | 12.50 | 6.25 | 6.25 | 6.25 |
| Spearman Correlations | +0.980 | +0.850 | +0.936 | +0.970 | - |

*Table 4: Percentile ranks by standardization mode, for scientists belonging to SDS Rural construction and environmental land management (AGR/10), for the publication period 2004-2008*

We repeat the preparation of rankings lists for each SDS of the Agricultural and veterinary science UDA, as seen in Table 5 (full SDS names are given in Annex). We observe that the rankings closest to benchmark are those obtained with $SS_{m0}$, with the Spearman coefficients oscillating between +0.970 and +0.999. The only exception is AGR/10 (as examined in Table 5), where the maximum correlation occurs under scenario $SS_{cit}$, whereby the number of observations is low (n=17). The lowest correlation value for the entire UDA ($\rho=+0.818$) is seen for Pedology (AGR/14), under scenario $SS_{perc}$: here too the number of observations is limited (n=18).

In general, for the Agricultural and veterinary sciences UDA, there is a high level of agreement between the rankings obtained under different scenarios and the ranking from the reference benchmark. To verify if this pattern extends to all the hard science UDAs, we prepare some descriptive statistics, giving values of correlation coefficients for each SDS of every UDA.

| SDS | n | $SS_{cit}$ | $SS_{perc}$ | $SS_a$ | $SS_{m0}$ | SDS | n | $SS_{cit}$ | $SS_{perc}$ | $SS_a$ | $SS_{m0}$ |
|---|---|---|---|---|---|---|---|---|---|---|---|
| AGR/02 | 95 | +0.949 | +0.951 | +0.976 | +0.994 | AGR/17 | 55 | +0.924 | +0.953 | +0.969 | +0.993 |
| AGR/03 | 92 | +0.947 | +0.920 | +0.954 | +0.996 | AGR/18 | 65 | +0.944 | +0.906 | +0.949 | +0.990 |
| AGR/04 | 25 | +0.948 | +0.971 | +0.972 | +0.996 | AGR/19 | 119 | +0.955 | +0.904 | +0.966 | +0.990 |
| AGR/05 | 29 | +0.971 | +0.969 | +0.976 | +0.996 | AGR/20 | 39 | +0.966 | +0.936 | +0.959 | +0.990 |
| AGR/07 | 76 | +0.951 | +0.957 | +0.984 | +0.995 | VET/01 | 88 | +0.963 | +0.968 | +0.990 | +0.998 |
| AGR/08 | 38 | +0.947 | +0.959 | +0.975 | +0.996 | VET/02 | 66 | +0.937 | +0.967 | +0.969 | +0.997 |
| AGR/09 | 38 | +0.955 | +0.888 | +0.975 | +0.994 | VET/03 | 72 | +0.961 | +0.940 | +0.960 | +0.996 |
| AGR/10 | 17 | +0.980 | +0.850 | +0.936 | +0.970 | VET/04 | 54 | +0.965 | +0.934 | +0.985 | +0.996 |
| AGR/11 | 69 | +0.977 | +0.962 | +0.981 | +0.996 | VET/05 | 73 | +0.964 | +0.953 | +0.980 | +0.996 |
| AGR/12 | 109 | +0.971 | +0.964 | +0.979 | +0.998 | VET/06 | 50 | +0.934 | +0.960 | +0.985 | +0.980 |
| AGR/13 | 103 | +0.969 | +0.962 | +0.986 | +0.997 | VET/07 | 35 | +0.973 | +0.981 | +0.978 | +0.996 |
| AGR/14 | 18 | +0.962 | +0.818 | +0.891 | +0.984 | VET/08 | 67 | +0.913 | +0.915 | +0.971 | +0.995 |
| AGR/15 | 140 | +0.963 | +0.961 | +0.982 | +0.998 | VET/09 | 47 | +0.931 | +0.931 | +0.970 | +0.994 |
| AGR/16 | 106 | +0.985 | +0.983 | +0.993 | +0.999 | VET/10 | 42 | +0.946 | +0.948 | +0.977 | +0.997 |

*Table 5: Spearman correlations between each scenario and the benchmark ($SS_{a0}$), for all SDSs of UDA Agricultural and veterinary sciences*

Under scenario 1 ($SS_{cit}$), the lowest average value of correlation is seen for two engineering UDAs, Civil engineering and Industrial and information engineering, both ρ=+0.942 (Table 6). However there are differences between the two UDAs: for Industrial and information engineering there is greater dispersion of correlation values (s.d.=0.067) due to the presence of an outlier SDS (Naval architecture, ING-IND/01), with a very limited number of faculty; Civil engineering has much more concentrated values of correlation (s.d. 0.026) around average and median values that almost coincide. The UDA with the highest average values of correlation is Chemistry (s.d. 0.008), where correlation values for the individual SDSs are concentrated in a narrow range between +0.958 and 0.985.

| UDA | Mean | St.dev. | Median | Minimum | Maximum |
|---|---|---|---|---|---|
| Mathematics and computer sciences | 0.961 | 0.016 | 0.965 | 0.938 | 0.988 |
| Physics | 0.954 | 0.047 | 0.977 | 0.864 | 0.986 |
| Chemistry | 0.979 | 0.008 | 0.982 | 0.958 | 0.985 |
| Earth sciences | 0.961 | 0.011 | 0.961 | 0.942 | 0.980 |
| Biology | 0.963 | 0.019 | 0.969 | 0.910 | 0.983 |
| Medicine | 0.960 | 0.032 | 0.968 | 0.800 | 0.986 |
| Agricultural and veterinary sciences | 0.955 | 0.017 | 0.958 | 0.913 | 0.985 |
| Civil engineering | 0.942 | 0.026 | 0.949 | 0.907 | 0.970 |
| Industrial and information engineering | 0.942 | 0.067 | 0.964 | 0.633 | 0.994 |

*Table 6: Descriptive statistics for values of correlation between the performance rankings calculated for each SDS using $SS_{cit}$ and using the benchmark $SS_{a0}$, by UDA*

Under scenario 2 ($SS_{perc}$), the average correlation values drop slightly but still remain above +0.900, and the UDAs that depart furthest from the benchmark ranking remain the same (Table 7). The SDSs with lowest correlation are again found in Industrial and information engineering (ρ=+0.652), and in Civil engineering (ρ= +0.754). However the SDS that had the lowest value in the previous scenario (Naval architecture) now has +1 correlation.

Finally, under scenarios 3 ($SS_a$) and 4 ($SS_{m0}$), there are very high average correlation values for the rankings of UDAs, respectively exceeding +0.950 and +0.980 (Table 8 and Table 9). Scenario $SS_{m0}$ is the one that generates the rankings closest to the benchmark scenario: no SDS drops below the +0.800 mark. Under all the scenarios, we observe that SDSs with low numbers of researchers are generally more sensitive to the type of standardization selected.

| UDA | Mean | St.dev. | Median | Minimum | Maximum |
|---|---|---|---|---|---|
| Mathematics and computer sciences | 0.936 | 0.021 | 0.941 | 0.888 | 0.960 |
| Physics | 0.952 | 0.018 | 0.951 | 0.923 | 0.976 |
| Chemistry | 0.968 | 0.009 | 0.971 | 0.946 | 0.977 |
| Earth sciences | 0.950 | 0.024 | 0.953 | 0.895 | 0.978 |
| Biology | 0.956 | 0.018 | 0.963 | 0.911 | 0.983 |
| Medicine | 0.957 | 0.023 | 0.962 | 0.859 | 1.000 |
| Agricultural and veterinary sciences | 0.940 | 0.038 | 0.953 | 0.818 | 0.983 |
| Civil engineering | 0.910 | 0.072 | 0.942 | 0.754 | 0.963 |
| Industrial and information engineering | 0.923 | 0.060 | 0.939 | 0.652 | 1.000 |

*Table 7: Descriptive statistics for values of correlation between the performance rankings calculated for each SDS using $SS_{perc}$ and using the benchmark $SS_{a0}$, by UDA*

| UDA | Mean | St.dev. | Median | Minimum | Maximum |
|---|---|---|---|---|---|
| Mathematics and computer sciences | 0.975 | 0.010 | 0.976 | 0.960 | 0.990 |
| Physics | 0.987 | 0.014 | 0.993 | 0.957 | 0.997 |
| Chemistry | 0.991 | 0.005 | 0.990 | 0.981 | 0.996 |
| Earth sciences | 0.981 | 0.010 | 0.985 | 0.958 | 0.989 |
| Biology | 0.989 | 0.006 | 0.992 | 0.977 | 0.996 |
| Medicine | 0.987 | 0.019 | 0.992 | 0.869 | 1.000 |
| Agricultural and veterinary sciences | 0.970 | 0.020 | 0.976 | 0.891 | 0.993 |
| Civil engineering | 0.955 | 0.024 | 0.965 | 0.920 | 0.981 |
| Industrial and information engineering | 0.960 | 0.040 | 0.965 | 0.738 | 1.000 |

*Table 8: Descriptive statistics for values of correlation between the performance rankings calculated for each SDS using $SS_a$ and using the benchmark $SS_{a0}$, by UDA*

| UDA | Mean | St.dev. | Median | Minimum | Maximum |
|---|---|---|---|---|---|
| Mathematics and computer sciences | 0.994 | 0.003 | 0.995 | 0.987 | 0.997 |
| Physics | 0.996 | 0.003 | 0.997 | 0.989 | 0.998 |
| Chemistry | 0.998 | 0.001 | 0.998 | 0.994 | 0.999 |
| Earth sciences | 0.996 | 0.002 | 0.997 | 0.993 | 0.998 |
| Biology | 0.996 | 0.003 | 0.997 | 0.986 | 0.999 |
| Medicine | 0.992 | 0.029 | 0.997 | 0.800 | 0.999 |
| Agricultural and veterinary sciences | 0.993 | 0.006 | 0.996 | 0.970 | 0.999 |
| Civil engineering | 0.995 | 0.003 | 0.997 | 0.990 | 0.998 |
| Industrial and information engineering | 0.988 | 0.017 | 0.995 | 0.941 | 1.000 |

*Table 9: Descriptive statistics for values of correlation between the performance rankings calculated for each SDS using $SS_{m0}$ and using the benchmark $SS_{a0}$, by UDA*

In light of these results, we could affirm that, generally, rankings of individuals obtained under different types of standardization do not show significant discrepancies with respect to rankings generated under the benchmark methodology. However, to obtain clearer resolution of the effects of various modes of standardization on rank of researchers, we further deepen the analysis at the individual level.

Beginning from the rankings by SDS under each standardization mode, we group researchers into four classes, as is common in research assessment exercises: the classes are the first to fourth quartiles for scientific strength, in each SDS. After preparing these classes under each scenario, we observe the researchers changing quartile with respect to the benchmark ranking list, calculated in percentages per UDA. Table 10 presents the results for each scenario. The highest values are registered under scenario $SS_{perc}$, with particularly notable observations of shift values in the UDAs of Mathematics and computer science (27.8%), Civil engineering (27.3%) and Earth sciences (26.7%). In these UDAS more than 26% of researchers would be classified differently from the benchmark. The lowest number of different cases of classification occurs under scenario $SS_{m0}$. This type of standardization causes its highest effects for the Earth sciences UDA, with approximately 8 out of every 100 researchers classified differently. The rankings for the Chemistry UDA are the least sensitive to the different modes of standardization.

| UDA | $SS_{cit}$ vs. $SS_{a0}$ | $SS_{perc}$ vs. $SS_{a0}$ | $SS_a$ vs. $SS_{a0}$ | $SS_{m0}$ vs. $SS_{a0}$ |
|---|---|---|---|---|
| Mathematics and computer sciences | 19.9% | 27.8% | 17.5% | 7.8% |
| Physics | 17.1% | 22.7% | 8.3% | 5.6% |
| Chemistry | 14.5% | 18.1% | 7.9% | 4.0% |
| Earth sciences | 22.3% | 26.7% | 13.6% | 8.1% |
| Biology | 18.6% | 20.7% | 9.7% | 5.6% |
| Medicine | 18.3% | 20.7% | 8.8% | 5.1% |
| Agricultural and veterinary sciences | 21.3% | 25.9% | 16.3% | 6.3% |
| Civil engineering | 24.8% | 27.3% | 16.5% | 6.4% |
| Industrial and information engineering | 18.9% | 25.9% | 19.1% | 5.7% |
| Total | 18.6% | 22.6% | 11.7% | 5.6% |

*Table 10: Percentage of researchers per UDA that change quartile rank when a standardization mode different from the benchmark is used*

We inquire further into shifts in rankings for top performers (1st quartile class) and bottom performers (4th quartile class). Table 11 presents the percentages of the top scientists in scenario $SS_{a0}$ that do not reach such classification under ranking lists from other standardization. In almost all UDAs, the failure to use standardization ($SS_{cit}$), or standardization by percentile of citations relative to the publications produced ($SS_{perc}$), causes jumps for quite high percentages of top performers. In an application of scenario $SS_{perc}$, 19.6% of top scientists in Mathematics and computer sciences would not be classified as top.

| UDA | $SS_{cit}$ vs. $SS_{a0}$ | $SS_{perc}$ vs. $SS_{a0}$ | $SS_a$ vs. $SS_{a0}$ | $SS_{m0}$ vs. $SS_{a0}$ |
|---|---|---|---|---|
| Mathematics and computer sciences | 13.6% | 19.6% | 12.1% | 4.8% |
| Physics | 12.5% | 17.3% | 5.1% | 4.3% |
| Chemistry | 8.8% | 12.5% | 5.5% | 2.5% |
| Earth sciences | 14.0% | 18.7% | 9.3% | 4.2% |
| Biology | 9.7% | 14.2% | 5.0% | 3.6% |
| Medicine | 9.8% | 13.3% | 6.0% | 3.2% |
| Agricultural and veterinary sciences | 12.1% | 14.8% | 11.1% | 3.5% |
| Civil engineering | 9.8% | 17.7% | 9.8% | 3.0% |
| Industrial and information engineering | 11.1% | 16.3% | 13.3% | 2.6% |
| Total | 10.7% | 15.0% | 7.7% | 3.4% |

*Table 11: Percentage of top scientists per UDA who do not result as top when a standardization mode different from the benchmark is used*

The Chemistry UDA shows the most contained levels of variation, across the different modes of standardization. Standardization to the median ($SS_{m0}$) provides the rankings lists with the most contained levels of variation across all the UDAs (never over 5%). The rankings lists applying the standardization to the citations average of all publications ($SS_a$) show strong variation among the UDAs, compared to the benchmark.

Table 12 presents the rates of variation due to application of standardizations different from the benchmark to the bottom class of performers. Among the alternatives, application of the median is again the least sensitive scaling factor, while this time notable differences occur for non-standardized citations.

| UDA | $SS_{cit}$ vs. $SS_{a0}$ | $SS_{perc}$ vs. $SS_{a0}$ | $SS_a$ vs. $SS_{a0}$ | $SS_{m0}$ vs. $SS_{a0}$ |
|---|---|---|---|---|
| Mathematics and computer sciences | 16.3% | 19.4% | 9.6% | 4.6% |
| Physics | 10.2% | 12.7% | 5.3% | 2.9% |
| Chemistry | 11.0% | 10.3% | 4.1% | 2.5% |
| Earth sciences | 19.1% | 14.0% | 7.4% | 7.9% |
| Biology | 12.7% | 11.9% | 6.0% | 3.2% |
| Medicine | 14.6% | 13.2% | 5.4% | 3.0% |
| Agricultural and veterinary sciences | 14.7% | 17.5% | 9.8% | 3.8% |
| Civil engineering | 23.8% | 15.2% | 10.4% | 6.7% |
| Industrial and information engineering | 16.0% | 16.9% | 11.5% | 5.9% |
| Total | 14.3% | 14.0% | 7.0% | 3.8% |

*Table 12: Percentage of bottom scientists per UDA who would not result as bottom when a standardization mode different from the benchmark is used*

## Conclusions

In bibliometrics, any serious comparative analysis of research performance for individuals or organizations requires field standardization of citations, due to the presence of different citations behaviors across different fields of research. A preceding study by the authors (Abramo et al., 2012) demonstrated that the citations average (with the average based only on publications actually cited) seems the most effective scaling factor. This work has followed up on the previous analysis, calculating the variation caused by the application of other scaling factors to individual researchers' productivity rankings. We developed different scenarios for producing individual performance rankings, one without any standardization and the other four applying the following scaling factors: percentile of citations, citations average and median of cited-only publications, and citation average of all publications.

We assumed the rankings derived from standardization to the citations average (for cited-only publications) as benchmark, and we then calculated the variations from rankings lists constructed using other standardization methods.

When we compared rankings from the different scenarios, the correlation values were quite high, with ρ values never dropping below +0.800. The scaling factor that produces the lowest rate of variation (highest coefficient of correlation) is the median citation of cited-only publications. Standardization to the percentile of citations received would produce the greatest variations. Fields with a low number of researchers obviously tend to experience greater rankings fluctuations between one method and another, due to the scarcity of data present. However these oscillations are always contained. The disciplines most subject to variations due to choice of a scaling factor different from the benchmark are engineering, agricultural and veterinary sciences, and mathematics and computer science. The remaining disciplines, particularly chemistry and physics, are less sensitive to the type of standardization used.

We also classified the rankings from the different scenarios into quartiles, to enable a further measure of the variation from the comparison to the benchmark scenario, in terms of percentage of faculty that shift classes. If no standardization is carried out, or if the percentile of citations is applied, the percentage of those that would not be classified in the benchmark's quartile results as around 20%. Variation is around 10% if the scaling factor chosen is the citations average of all publications, and 5% if standardization is to the median. For the classification of the top 25% and the bottom 25% of researchers, the scaling factor that causes the greatest variation differs but the one that introduces the least variation, in both cases, is the same.

Ultimately, the scaling factor that generally produces the least variation is the citations median of cited-only publications. However this variation is only "least" among the various standardization methods that are compared to the benchmark scaling factor: the citation average of cited-only publications.

**Annex – list of SDS in the Agricultural and veterinary sciences UDA**

| Code | Title |
|---|---|
| AGR/02 | Agronomy and Herbaceous Cultivation |
| AGR/03 | General Arboriculture and Tree Cultivation |
| AGR/04 | Horticulture and Floriculture |
| AGR/05 | Forestry and Silviculture |
| AGR/07 | Agrarian Genetics |
| AGR/08 | Agrarian Hydraulics and Hydraulic Forest Management |
| AGR/09 | Agricultural Mechanics |
| AGR/10 | Rural construction and environmental land management |
| AGR/11 | General and Applied Entomology |
| AGR/12 | Plant Pathology |
| AGR/13 | Agricultural Chemistry |
| AGR/14 | Pedology |
| AGR/15 | Food Sciences |
| AGR/16 | Agricultural Microbiology |
| AGR/17 | General Techniques for Zoology and Genetic Improvement |
| AGR/18 | Animal Nutrition and Feeding |
| AGR/19 | Special Techniques for Zoology |
| AGR/20 | Animal Husbandry |
| VET/01 | Anatomy of Domestic Animals |
| VET/02 | Veterinary Physiology |
| VET/03 | General Pathology and Veterinary Pathological Anatomy |
| VET/04 | Inspection of Food Products of Animal Origin |
| VET/05 | Infectious Diseases of Domestic Animals |
| VET/06 | Parasitology and Parasitic Animal Diseases |
| VET/07 | Veterinary Pharmacology and Toxicology |
| VET/08 | Clinical Veterinary Medicine |
| VET/09 | Clinical Veterinary Surgery |
| VET/10 | Clinical Veterinary Obstetrics and Gynaecology |